\documentstyle[12pt]{article}

\begin{document}
\setlength{\parskip}{5mm}

\title{The Genuine Cosmic Rosetta}
\author{Robert V. Gentry and David W. Gentry
      \\The Orion Foundation
      \\P.O. Box 12067
      \\Knoxville, TN 37912
       }
\date{gr-qc/9806061}

\maketitle

\begin{abstract}

Reexamination of general relativistic experimental results shows the
universe is governed by Einstein's static-spacetime general relativity
instead of Friedmann-Lemaitre expanding-spacetime general relativity. The
absence of expansion redshifts in a static-spacetime universe suggests a
reevaluation of the present cosmology is needed.

\end{abstract}

For many decades the Friedmann-Lemaitre spacetime expansion redshift
hypothesis$^{1,2}$ has been accepted as the Rosetta of modern cosmology. It
is believed to unlock the mysteries of the cosmos just as the archaeological
Rosetta unlocked the mysteries of ancient Egypt. But are expansion redshifts 
{\it The Genuine Cosmic Rosetta}? Until now this has been the consensus
because of their apparent, most impressive ability to uniquely explain how
the twentieth century's two great astronomical and astrophysical
discoveries---meaning of course the Hubble redshift relation and the 2.7K
Cosmic Blackbody Radiation (CBR)---can be accounted for within the framework
of a hot big bang universe. But this consensus is not universal. For
example, Burbidge$^3$ and Arp$^4$ continue to note that while most
astronomers and astrophysicists accept the hot big bang and attribute
extragalactic redshifts to expansion effects, they continue to ignore the
minority view that certain observations, such as anomalous quasar redshifts,
imply the need for a different redshift interpretation, and perhaps a
different universe model as well.

What is now almost certain to attract more attention to the Burbidge/Arp
claim is the surprising, very recent discovery of a new redshift
interpretation$^5$ of the Hubble relation and the 2.7K CBR based on a
universe governed by Einstein's static-spacetime general relativity. This
discovery shows for the first time that the expansion redshift hypothesis is
not the only possible explanation of extragalactic redshifts. And in so
doing it inevitably focuses attention on the question of how the universe is
formatted, relativistically speaking: Is it governed by Friedmann-Lemaitre
expanding spacetime general relativity, as has been generally assumed for
many decades? Or does the new redshift discovery point instead to it being
governed by Einstein static-spacetime general relativity? There are three
solid reasons why this question should now be further investigated.

First, G. F. R. Ellis, one of the big bang's ablest advocates has: (i) gone
so far as to suggest the big bang might not be correct, (ii) cautioned
against the bandwagon effect in supporting it, (iii) emphasized the constant
need to question and test its {\it foundations}, and (iv) even entertained
the possibility of a paradigm shift away from it.$^6$ Is Ellis aware of
something that has eluded everyone else? Not really. Rather, his forthright
appraisal relates to the fact that the expanding spacetime paradigm stands
alone among all the theories of modern physics in that, even after many
decades, no way has yet been found to experimentally confirm the existence
of the cosmic expansion factor, $\Re $, which is the essential parameter in
Friedmann-Lemaitre expansion redshift equation, $z_{\exp }=\Re /\Re _e-1$.
Thus, despite the fact that expansion redshifts have been widely inferred to
exist because of their apparently successful use in uniquely accounting for
the Hubble redshift relation and the 2.7K CBR, we must recognize that
inference is not the same as certainty obtained by direct experimentation.
We should also recognize that the recent discovery of the new redshift
interpretation,$^5$ which shows the uniqueness part of the inference
argument has always been ill-founded, makes it more imperative than ever to
further probe the expanding spacetime paradigm.

In doing this we almost immediately come face-to-face with a most
interesting feature---namely, in defiance of long-established protocol for
testing any and all modern scientific theories for consistency with known
physical laws, we find wavelength expansion effects, which are the presumed
cause of expansion redshifts, have been authoritatively defined to be exempt
from obeying conservation of energy. For example, in 1981, 1989, 1990, and
1993, respectively, cosmologists Harrison,$^7$ Silk,$^8$ Alpher and
Herman,$^9$ and Peebles$^{10}$, independently concurred that the in-flight
photon
energy loss which accompanies photon wavelength expansion represents
nonconservation of energy. In 1993 Peebles stated the situation rather
plainly:

{\it ``However, since the volume of the universe varies as $a(t)^3$,
the net radiation energy in a closed universe decreases as $1/a(t)$ as the
universe expands. Where does the lost energy go? ... The resolution of this
apparent paradox is that while energy conservation is a good local concept,
....there is not a general global energy conservation law in general
relativity theory''.}$^{10}$

This conclusion is based on Peebles' use of the expanding-spacetime
paradigm. Even though such conclusions have long remained unchallenged, we
are unable to find where the full implications of this and similar assertions%
$^{7-10}$ have ever been critically analyzed and reported in a text or
journal. Indeed, we cannot even find where the answer to the most basic
question about how much radiation energy is predicted to have been lost due to
expansion effects has ever appeared in a journal publication. So we
undertake to do this now, and the answer is quite large. Consider in
particular the magnitude of the nonconservation-of-energy loss of CBR
photons as in theory they were expansion-redshifted from 3000K at decoupling
to the present 2.7K.

Assuming a nominal universe volume, $V_{univ}$, of 15 billion ly radius, the
2.7K CBR having about $\overline{n}=410$ photons-cm$^{-3}$ with average
energy of about $\overline{\varepsilon }_{2.7}=10^{-15}$ erg, and the 3000K
radiation with $\overline{\varepsilon }_{3000}=1.13\times 10^{-12}$ erg,
and an equal
number of photons,$^8$ we compute the total CBR expansion energy loss as $%
E_{\exp }=\overline{n}\times (\overline{\varepsilon }_{3000}-\overline{%
\varepsilon }_{2.7})\times V_{univ}=5.5\times 10^{75}$ erg. This is
about three times the galactic mass of a universe composed of 10$^{21}$
solar masses. For an initial fireball temperature of 3 million K, the total
radiation energy loss would be three thousand times the mass of such a
universe. Even more incredibly, since in theory photon conservation$^8$
extends back to a fireball temperature of 30 billion K, in this case the
theorized nonconservation-of-energy loss projects to be thirty million times
the mass of such a universe.

These gargantuan energy losses command our attention for there appear to be
only two ways to interpret them, and both have significant cosmological
implications. If expanding spacetime general relativity and expansion
redshifts correctly describe the universe we inhabit, it would seem that our
long-held concepts of energy conservation are drastically in error. On the
other hand, if we hold to universal energy conservation, then it would seem
our universe must be governed by Einstein's static-spacetime general
relativity and Einstein redshifts, which are consistent with energy
conservation. As this Letter now reports, even though the experimental data
needed to distinguish these alternatives have existed for more than two
decades, their cosmological implications have remained virtually unnoticed
until now.

Testing the expanding-spacetime universe paradigm begins with listing its
twofold basic assumption---namely, that general relativistic processes
operate to expand wavelengths only while photons are in-flight. It is
imperative to assume complete cessation of expansion effects during
emission/absorption in order to insure agreement with the astronomical
requirement of a fixed emission wavelength, $\lambda _e.$ However, when we
examine the many relativistic gravitational experiments performed over the
last few decades we find that, while those results conflict with the
expansion paradigm's basic assumptions, they are completely in accord with
the predictions of the static-spacetime theory of general relativity as
Einstein first proposed it in 1916.$^{11}$

In that seminal paper he predicted that gravity should cause a perfect clock
to go ``{\it ... more slowly if set up in the neighborhood of ponderable
masses. From this it follows that the spectral lines of light reaching us
from the surface of large stars must appear displaced towards the red end of
the spectrum.''}$^{11}$

In 1954 Brault's redshift measurement$^{12}$ of the sodium D line
emanating from the sun's spectrum did succeed in confirming the {\it
magnitude%
} of the gravitational redshift that Einstein had predicted. But this result
did not settle the question of its {\it origin}.  More specifically,
was Einstein correct in postulating that different gravitational potentials
at source and observer meant that clocks at these locations should run at
intrinsically different rates, and hence that this was the origin of the
gravitational redshift? Or did the measured redshift instead have its origin
in photons experiencing an in-flight energy exchange with gravity as they
moved in a changing gravitational potential in their transit from a star to
the Earth?

Even the 1965 Pound-Snider experiments$^{13}$ did not settle this question.
True, these observers did find a $\Delta \nu /\nu =-\Delta \varphi
/c^2=gh/c^2$ fractional frequency difference between $^{57}$Fe gammas
emitted at the top and received at the bottom of a tower of height, $h$, and
this result did more precisely confirm the {\it magnitude} of the Einstein
redshift. But it did not settle its {\it origin}, for they could not tell
whether the redshift resulted from in-flight wavelength change as the photon
passed through a gravitational gradient, or whether it was due instead to
differences in gravity affecting the relative frequency at the point of
emission. They did suggest, however, this issue could be decided by
comparing coherent light sources operating at different potentials.$^{13}$

That is, if atomic clocks separated by a height $h$ were found to run at the
same rate, this would prove that local gravity does not affect relative
emission frequencies, and hence that relativistic redshifts do result from
photons experiencing an in-flight energy exchange with gravity. If this had
been the experimental outcome, then the predictions of the
expanding-spacetime paradigm, with its expansion redshifts, would have been
fully confirmed.

But as is now well-known, atomic clock experiments have repeatedly shown
that a clock on a mountain top does run faster than its sea level
counterpart by a fractional amount $\Delta \nu /\nu =-\Delta \varphi /c^2=$
$gh/c^2$, the same shift found by Pound and Snider. Although not generally
recognized as such until now, this result proved long ago that the Einstein
redshift is due to local gravity operating to affect relative emission
frequencies as seen by an observer in a different gravitational potential.
Moreover, the basic principle of local gravity affecting relative emission
frequencies is further confirmed many thousands of times every hour in the
continuing operation of GPS atomic clocks. Synchronization of those clocks
utilizes the Einstein static-spacetime paradigm with its predicted effect of
gravity on emission frequency to calculate how much faster satellite clocks
will be expected to operate once they are in orbit. Thus, prior to launch,
satellite clocks are preset to run about 38,400 ns/d slower than the base
master clock to compensate for their faster rate in orbit.$^{14}$

Another remarkable confirmation of gravity's effect on emission frequencies
comes from Taylor's comparison of atomic clock time with pulsar timing data.$%
^{15}$ To synchronize both data sets he found it necessary to account for
the change of local atomic clock time due to the monthly variation in the
sun's gravitational potential at Earth. In Taylor's own words, {\it ``Here
is direct proof, based on a clock some 15,000 light years from the solar
system, that clocks on Earth run more slowly when the moon is full---because
at this time of the month we are deeper in the gravitational potential of
the sun!''}$^{15}$

Thus Einstein's 1916 predictions about both the origin and the magnitude of
the gravitational redshift have been confirmed by a variety of general
relativistic experiments, so as to obtain the following conclusions: (1)
there is only one gravitational redshift between two points at different
potentials, and it is given by $\Delta \nu /\nu
=-\Delta \lambda /\lambda =-\Delta \varphi /c^2$, and (2) this redshift does
not originate with photons exchanging energy with gravity during transit
through a potential gradient, but instead originates in precisely the way
that Einstein stated it in 1916, and again in 1952---namely, {\it ``An atom
absorbs or emits light of a frequency which is dependent on the potential of
the gravitational field in which it is situated.''}$^{16}$

The foregoing results contradict the basic assumptions of a universe
governed by Friedmann-Lemaitre expanding-spacetime general relativity,
showing instead that the universe we inhabit is one governed by Einstein's
static-spacetime general relativity. In doing this they focus added
attention on the recent discovery of {\it a New Redshift Interpretation}$^5$
(NRI)---which shows for the first time that an expanding universe
characterized by Hubble-relation galactic recession and the 2.7K CBR can be
explained within the framework of a universe governed by static-spacetime
general relativity. The credibility of the NRI is enhanced by its apparent
ability to also account for:$^5$ (i) the 2.7K CBR's spatial 
isotropy, (ii) the predicted variation of redshift, $z$, with CBR
temperature, (iii) the observed monotonic decline in galactic angular size
with increasingly higher redshifts, and (iv) possibly the sparsity of high
redshift quasars for $z>4$.

Thus this Letter concludes that Einstein's static-spacetime general
relativity is indeed {\it The Genuine Cosmic Rosetta. }Its apparent success
in interpreting the aforementioned observations implies it now needs to be
further tested against an increasing array of other astrophysical phenomena.
Indeed, in the relatively short time that has elapsed since the NRI's
publication, new results have appeared which seem to provide one of the
strongest observational tests of its validity. We refer to most recent
reports of astronomical observations strongly suggesting the existence of a
repulsive force in the outermost reaches of the universe.$^{17,18}$ An
important question which may soon attract wide attention is whether these
observations may reasonably be interpreted to be a remarkable confirmation
of the NRI's prediction that ours is a universe dominated by a repulsive
force due to vacuum gravity.$^5$

In another paper we show how the NRI and a static-spacetime universe lead
to new possibilities for quasar redshifts.$^{19}$ The latter may be of
considerable interest to researchers such Burbidge and Arp, who have long
contended that certain quasars provide strong evidence of intrinsic
redshifts.  Also, while we acknowledge the concerns and results of
Burbidge,$^3$ Arp,$^4$ Ellis,$^{6,20}$ and Ellis {\it et al.},$^{21}$ as
providing motivation for pursuing the investigation of this most
interesting topic, we do not imply that these researchers have been
participants in it.  

Where the results of this Letter may attract the most interest is with the
majority of astronomers and astrophysicists who have long believed the
creation of the universe can be traced to a big bang singularity, for the
results presented herein challenge the very existence of the big bang's
essential ingredient of spacetime expansion. These results are presented in
the spirit of free scientific inquiry with the expectation that more details
about these matters will emerge as all their ramifications are openly and
freely pursued.


\begin{thebibliography}{00}

\bibitem{1} A. Friedmann, {\it Z. Phys.} {\bf 10} (1922) 377.

\bibitem{2} G. Lemaitre, {\it Annales Soci\'et\'e Scientifique Bruxelles}
{\bf A47} (1927) 49.

\bibitem{3} G. Burbidge, {\it Ap. \& Space Science} {\bf 244} (1996) 169.

\bibitem{4} H. Arp, {\it Ap. \& Space Science} {\bf 244} (1996) 1.

\bibitem{5} R. V. Gentry, {\it Mod. Phys. Lett. A} {\bf 12} (1997) 2919;
also available at http://www.wspc.com.sg/ .

\bibitem{6} G. F. R. Ellis, ``Innovation, resistance and change: the
transition to
the expanding universe,'' in {\it Modern Cosmology in Retrospect} (Cambridge
Univ. Press, 1990), pp. 97--113.

\bibitem{7} E. R. Harrison, {\it Cosmology: Science of the Universe}
(Cambridge Univ.
Press, 1981), pp. 275--276.

\bibitem{8} J. Silk, {\it The Big Bang} (W. H. Freeman \& Co., 1989), pp.
423--425.

\bibitem{9} R. A. Alpher and R. Herman, ``Early work on big bang cosmology and
the CBR,'' in {\it Modern Cosmology in Retrospect} (Cambridge Univ. Press,
1990), pp. 151--152.

\bibitem{10} P. J. E. Peebles, {\it Principles of Physical Cosmology}
(Princeton
Univ. Press, 1993), pp. 96--99; 138--139.

\bibitem{11} A. Einstein, {\it Ann. der Physik} {\bf 49} (1916) 769. English
reprint in {\it The Principle of Relativity} (Dover Publications), pp.
160--164.

\bibitem{12} J. W. Brault, Abstract, ``Gravitational Redshift of Solar
Lines,'' in
{\it Bull. Amer. Phys. Soc.} {\bf 8} (1963) 28.

\bibitem{13} R. V. Pound and J. L. Snider, {\it Phys. Rev. B} {\bf 140}
(1965) 788.

\bibitem{14} N. Ashby and J. J. Spilker, Jr. in {\it The Global
Positioning System: Theory and Applications., Vol. 1} (American Institute of
Aeronautics and Astronautics, Inc., 1995), Chapter 18.

\bibitem{15} J. H. Taylor, ``Astronomical and Space Experiments To Test
Relativity,''
in {\it General Relativity and Gravitation} (Cambridge Univ. Press, 1987), p.
214.

\bibitem{16} A. Einstein, {\it Relativity: The Special and General Theory}
(Crown
Trade Paperbacks, New York, 1961), p. 130.

\bibitem{17} S. Perlmutter {\it et al.}, LBNL preprint 41801 (1998);
see also www-supernova.lbl.gov .

\bibitem{18} A.G. Riess {\it et al.}, astro-ph/9805201.

\bibitem{19} R. V. Gentry, in preparation.

\bibitem{20} G. F. R. Ellis, {\it Ann. Rev. Ast. Ap.} {\bf 22} (1984) 157.

\bibitem {21} G. F. R. Ellis, R. Maartens, and S. D. Nel, {\it Mon. Not.
Roy. Ast. Soc.} {\bf 184} (1978) 439. 
\end{thebibliography}
\end{document}